\documentclass[aps,prl,floatfix,reprint,a4paper,showpacs,showkeys,nobibnotes]{revtex4-1}
\usepackage[utf8x]{inputenc}
\usepackage[english]{babel}
\usepackage{ucs}
\usepackage{amsmath}
\usepackage{amsfonts}
\usepackage{amssymb}
\usepackage{graphicx}
\usepackage{dcolumn}
\usepackage{bm}

\begin{document}
\title{Stochastic method for in-situ damage analysis}
\author{Philip Rinn}
\email{philip.rinn@uni-oldenburg.de}
\author{Hendrik Hei{\ss}elmann}
\author{Matthias W\"achter}
\author{Joachim Peinke}
\email{peinke@uni-oldenburg.de}
\affiliation{ForWind -- Center for Wind Energy Research, Institute of Physics, Carl-von-Ossietzky University Oldenburg, 26111 Oldenburg, Germany}
\date{November 2, 2012}

\begin{abstract}
Based on the physics of stochastic processes we present a new approach for structural health monitoring. We show that the new method allows for an in-situ analysis of the elastic features of a mechanical structure even for realistic excitations with correlated noise as it appears in real-world situations. In particular an experimental set-up of undamaged and damaged beam structures was exposed to a noisy excitation under turbulent wind conditions. The method of reconstructing stochastic equations from measured data has been extended to realistic noisy excitations like those given here. In our analysis the deterministic part is separated from the stochastic dynamics of the system and we show that the slope of the deterministic part, which is linked to mechanical features of the material, changes sensitively with increasing damage. The results are more significant than corresponding changes in eigenfrequencies, as commonly used for structural health monitoring.
\end{abstract}


\maketitle

\section{Introduction}
It is a crucial task to achieve an early and reliable detection of feature-changes of mechanical structures caused by damage, fatigue, or other environmental influences. Commonly, those detection systems use fast Fourier transformation (FFT) to extract system features \cite{Carden2004,Hameed2009,Lu2009} and to determine the condition of the system from changes in the eigenfrequencies. One drawback of this approach is that noisy excitation of the structure broadens the peaks of the frequency spectrum and thus makes it harder to detect changes reliably. In this paper we present a method to obtain the dynamical behavior of the system and to analyze changes of the system's dynamics due to damages. Our proposed method, which is based on the physics of stochastic processes, is suitable for in-situ application, as we show that it is robust against changing working conditions. In particular, we show how to separate the stochastic response dynamics of the system from the deterministic one. Even for different levels of exciting noise or, respectively, of turbulence we are able to analyze only the determinism of the system dynamics directly linked to the mechanical properties of the system.

The paper is organized as follows. First we present the selected experimental system and its numerical model. In section three the method and the results are presented.

\section{Experimental system}
\begin{figure}
  \centering
  \includegraphics{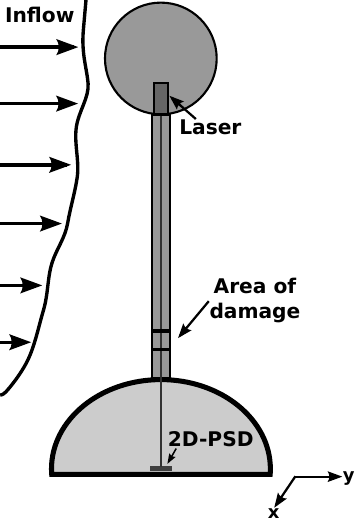}
  \caption{\label{schema} Schematic of the beam structure.}
\end{figure}

As a simple mechanical system we used a one-sided fixed beam structure and placed it in the wind tunnel of the University of Oldenburg (cf. Fig.~\ref{schema}). To investigate a turbulent flow-structure interaction, a motor-driven gust generator was used for the production of the turbulent inflow conditions. A sphere was mounted on top of the beam to increase the acting drag force. The deflection of the structure in the horizontal $xy$-plane was measured with a laser diode mounted in the sphere and aimed onto a two-dimensional position sensitive detector (2D-PSD). This light pointer principle is known to be highly resolving, it is used in atomic force microscopy and has recently been used also for new anemometers \cite{Puczylowski2010,Heisselmann2010}. Thus, the bending of the beam in the range of $\mu$m can be resolved within $\mu$sec.

\begin{figure}
  \centering
  \includegraphics{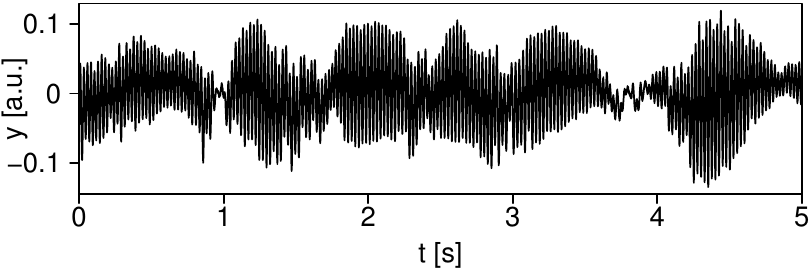}
  \caption{\label{time_series} Time series of the deflection in $y$-direction.}
\end{figure}

For our measurements two steel beams were prepared. While one beam remained undamaged, the dynamical behavior of the other beam was changed in two steps. In the first step the beam was treated thermally, it was heated at red heat  and cooled down fast and in the second step the beam was cut at a length of 40$\,$\% of its circumference. For four different wind speeds ($7\,$m/s, $10\,$m/s, $15\,$m/s and $17\,$m/s) time series of ten minutes length were recorded with a sampling frequency of $30\,$kHz. From the measured data the deflection of the beam in $x$- and $y$-direction was calculated. Figure~\ref{time_series} presents a segment of the measured time series of the deflection in $y$-direction.

In addition to the experiment a numerical model of the system has been set up. The beam structure was modeled by a finite element model in which four elements were used, three for the beam and one for the sphere. The mechanical properties were chosen in accordance with the parameters of the experimental set-up. To simulate the damage the stiffness of the second element was reduced in several steps up to $70\,$\%. The acting forces were calculated from a series of Gaussian distributed white noise ($\bar{u}_x = 7$, $\bar{u}_y = 13$, $\sigma_{x,y} = 2$). The forces were fed into the model at the nodes, the distribution was calculated according to the geometry and the drag coefficients. Time series of ten minutes length where generated at a sampling frequency of $30\,$kHz in order to match the parameters of the experiments.

\section{Stochastic Analysis}
A wide range of dynamic systems (in particular if fluctuating, noisy forces are involved) can be described by stochastic differential equations, namely, the Langevin equation \cite{Risken1996}

\begin{equation}
\dot{X}_i(t) = D^{(1)}_i(X(t)) + \sum^2_{j=1} [\sqrt{D^{(2)}(X(t))}]_{ij}\,\Gamma_j(t) \, \mathrm{.}
\label{LangevinD}
\end{equation}

The time derivative of the system variable $\dot{X}(t)$ ($X \in \mathbb{R}^n$) can be expressed as a sum of a deterministic part $D^{(1)}$ and the product of a stochastic force $\Gamma(t)$ and a weight coefficient $D^{(2)}$. For an ideal process the stochastic force $\Gamma(t)$ is white noise with zero mean, i.e. it is $\delta$-correlated and Gaussian distributed. In Eq.~(\ref{LangevinD}) the symbol $[\cdot]_{ij}$ refers to the element $(i,j)$ of the resulting matrix. Throughout the paper, we apply Itô's interpretation of stochastic integrals (cf.~\cite{Risken1996}).

To show how our experimental situation can be linked to such a stochastic differential equation, we start with the idea that the system variable $X(t)$ is the deflection. As we are interested in the dynamics we propose that for $X(t)$ a general differential equation

\begin{equation}
\dot{X} = f(X(t), u(t))
\end{equation}

holds. Here $f$ denotes an unknown function characterizing the dynamics, which depends on the deflection $X(t)$ and the wind velocity $u(t)$. The turbulent wind velocity acting on the structure can be split up into the sum of the mean wind speed $\bar{u}$ and its fluctuations $u'(t)$

\begin{equation}
u(t) = \bar{u} + u'(t) \; \; \text{with} \; \langle u'(t) \rangle = 0 \, \mathrm{.}
\label{windspeed}
\end{equation}

The temporal development of $X(t)$ is obtained by integration

\begin{equation}
X(t+\tau) - X(t) = \int_{t}^{t+\tau} f(X, u(t)) \, \mathrm{d}t \, \mathrm{.}
\label{xtau}
\end{equation}

Using the linearization of $f(X,u)$ with respect to $u$

\begin{equation}
f(X,u) = f(X,\bar{u}) + u'(t) \left[ \frac{\partial f(X,\bar{u})}{\partial u} + \frac{\partial f(X,\bar{u})}{\partial X}\frac{\mathrm{d}X}{\mathrm{d}u} \right]
\end{equation}

and that $f(X,\bar{u})$, $\frac{\partial f}{\partial u}$ and $\frac{\partial f}{\partial X}\frac{\mathrm{d}X}{\mathrm{d}u}$ are slowly varying for small $\tau$, Eq.~(\ref{xtau}) can be written as

\begin{equation}
X(t+\tau) - X(t) = \tau \cdot f(X,\bar{u}) + \left[ \frac{\partial f}{\partial u} + \frac{\partial f}{\partial X}\frac{\mathrm{d}X}{\mathrm{d}u} \right] \int_{t}^{t+\tau} u'(t) \, \mathrm{d}t \, \mathrm{.}
\end{equation}

Analyzing experimental data, the mean value of $\langle X(t+\tau) - X(t) \rangle |_{X(t) = \bm{x} }$ can be estimated by taking all $X(t)$ values which are in a close neighborhood to a chosen value $\bm{x}$. This conditional mean is

\begin{equation}
\langle X(t+\tau) - X(t) \rangle |_{X(t) = \bm{x}} = \tau \cdot f(\bm{x},\bar{u}) \, \mathrm{,}
\label{CondMom}
\end{equation}

using

\begin{equation*}
\langle \tau \cdot f(X,\bar{u}) \rangle |_{X(t) = \bm{x}} = \tau \cdot f(\bm{x},\bar{u})
\end{equation*}

and, using $\langle u'(t) \rangle |_{X(t) = \bm{x}} = 0$ according to Eq.~(\ref{windspeed}),

\begin{eqnarray*}
\left[ \frac{\partial f(\bm{x} ,\bar{u})} {\partial u} + \frac{\partial f(\bm{x} ,\bar{u})}{\partial X}\frac{\mathrm{d}X}{\mathrm{d}u} \right] \left.\left\langle \int_{t}^{t+\tau} u'(t) \, \mathrm{d}t \right\rangle \right|_{X(t) = \bm{x} } &=& \\
\left[ \frac{\partial f(\bm{x} ,\bar{u})} {\partial u} + \frac{\partial f(\bm{x} ,\bar{u})}{\partial X}\frac{\mathrm{d}X}{\mathrm{d}u} \right] \int_{t}^{t+\tau} \langle u'(t) \rangle |_{X(t) = \bm{x} } \, \mathrm{d}t &=& 0 \, \mathrm{.}
\end{eqnarray*}

The connection to the Langevin equation (\ref{LangevinD}) can be seen by expressing the drift and diffusion terms as Kramers-Moyal coefficients \cite{Risken1996}, which are the values of the conditional moments for $\tau \rightarrow 0$

\begin{eqnarray}
D^{(1)}_i(x) &=& \lim_{\tau \rightarrow 0} \frac{1}{\tau} \langle (X_i(t+\tau) - x_i) \rangle |_{X(t) = \bm{x}} \label{Dn}\\
D^{(2)}_{ij}(x) &=& \lim_{\tau \rightarrow 0} \frac{1}{2\tau} \langle (X_i(t+\tau) - x_i)(X_j(t+\tau) - x_j) \rangle |_{X(t) = \bm{x}} \, \mathrm{.}\nonumber
\end{eqnarray}

Siegert et al. \cite{Siegert1998} and Friedrich et al. \cite{Friedrich2000} developed a method to reconstruct drift $D^{(1)}$ and diffusion $D^{(2)}$ (Eq.~(\ref{Dn})) directly from measured data for stationary continuous Markov processes (for further details see also \cite{Friedrich2011}).

Equation~(\ref{LangevinD}) should be interpreted in the way that for every time $t$ where the system meets an arbitrary but fixed point $\bm{x}$ in phase space, $X(t+\tau)$ is determined by the deterministic function $D^{(1)}(x)$ and the stochastic function $\sqrt{D^{(2)}(x)}\Gamma(t)$. Both, $D^{(1)}(x)$ and $D^{(2)}(x)$, are constant for fixed $x$.

To analyze the dynamical behavior of our set-up only the deterministic part of the Langevin equation is needed. From the derivative of Eq.~(\ref{CondMom}) we see that the function $f(\bm{x},\bar{u})$ corresponds to $D^{(1)}$ and, more important for practical purpose, it is only essential to require that the mean of the fluctuations will vanish. This is a much weaker requirement than requiring delta correlated and Gaussian distributed noise.

\begin{figure*}
  \centering
  \includegraphics{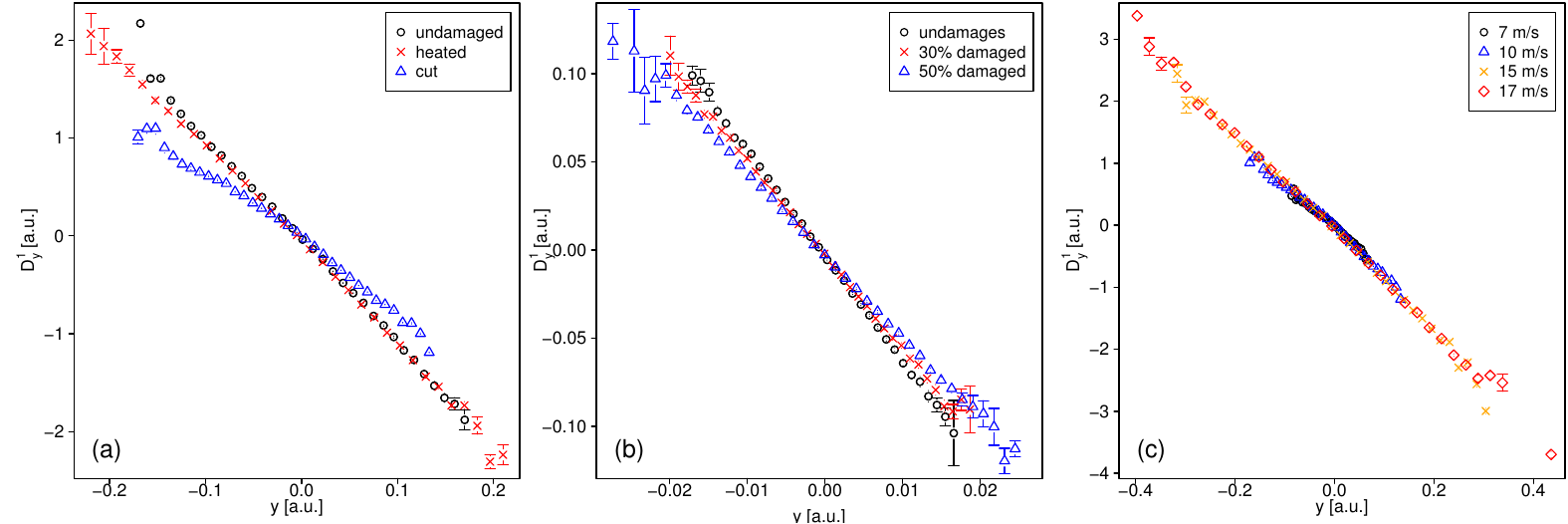}
  \caption{\label{slope} \label{FEM} \label{speeds} (Color online) Drift function in $y$-direction conditioned on $x = 0$ (a) for three damages at $\bar{u} = 10\,$m/s (measurement), (b) for three damages (simulation) and (c) of the cut beam for different wind speeds.}
\end{figure*}

For our measured data we proceed as follows. The $x$- and $y$-coordinates of the deflection of the beam span the phase space of $X(t)$ which was divided into 40 equidistant bins in each direction (fixing different $\bm{x}$-values). The drift function was calculated with Eq.~(\ref{Dn}) point-wise for each bin resulting in two $40\times40$ matrices for $x$- and $y$-direction. Figure~\ref{slope}a shows a cut in $y$-direction through the drift function for the main flow direction ($y$-direction) at $\bar{u} = 10\,$m/s. The slope of the heated beam is $6.3\% \pm 0.7\%$ smaller than of the undamaged one while the slope of the cut beam is $28.2\% \pm 1.3\%$ smaller. Larger errors in the outermost bins are due to the small number of events in these bins as large deflections are not as frequent as small ones. Table \ref{tabProzSlope} shows the change of the slopes of the drift function for all wind speeds. The values are presented in percental values normalized to the undamaged values for each wind speed.

\begin{table}
\caption{\label{tabProzSlope} Percental values of the slope of the drift function in $y$-direction conditioned on $x = 0$ (normalized to the undamaged values for each wind speed).}
\begin{ruledtabular}
\begin{tabular}{rccc}
  & undamaged & heated & cut \\ \hline
 7 m/s & $100.0\% \pm 1.2\%$ & $94.3\% \pm 1.4\%$ & $74.1\% \pm 1.5\%$ \\ 
 10 m/s & $100.0\% \pm 0.4\%$ & $93.7\% \pm 0.7\%$ & $71.8\% \pm 1.3\%$ \\ 
 15 m/s & $100.0\% \pm 1.0\%$ & $96.2\% \pm 1.5\%$ & $71.8\% \pm 0.8\%$ \\ 
 17 m/s & $100.0\% \pm 1.8\%$ & $97.3\% \pm 1.9\%$ & $68.6\% \pm 0.6\%$ \\ 
\end{tabular}
\end{ruledtabular}
\end{table}

From the numerical model simulations a similar behavior was found. Figure~\ref{FEM}b shows that an increasing reduction of the stiffness leads to a decrease of the slope of the drift. For a reduction of the stiffness by $30\%$ the slope decreases by $11.6\% \pm 0.8\%$, for a reduction of $50\%$ the decrease is $22.9\% \pm 0.4\%$.

The change of the slope can be made plausible when one links the drift to mechanical features of the material and interprets it as an indicator how fast the beam returns to its position of rest. Here it might be noted that the negative slope of $D^{(1)}$ corresponds to an attraction to a position of rest defined by the fixed point $D^{(1)}(x=0,y) = 0$. A decreasing restoring force then results in a decreasing slope of the drift. The cut in the beam means a major decrease of its stiffness thus the slope of the drift should be significantly smaller. ($D^{(1)}$ is asymmetric for $y<0$ and $y>0$, see Fig.~\ref{slope}a, this is likely to be due to the fact that a cut in a beam leads to asymmetric weakening of the material, whereas in the numerical model the weakening in the volume element was symmetric, compare Fig.~\ref{FEM}b.)

As the mean wind speed is not constant in free field conditions it is important to know if the slope of the drift changes with respect to the wind speed, or more generally speaking how $D^{(1)}$ depends on $\bar{u}$. This is particularly important for our case where we only condition on bins in $\bm{x}$ and thus the conditional moments of Eq.~(\ref{Dn}) sum over different $\bar{u}$. The obtained results are shown in Fig.~\ref{speeds}c and in Table~\ref{tabProzSlope}. We conclude that independent of a questionable or weak dependence of the slope of $D^{(1)}$ on wind speed the effect of the damages can be clearly quantified.

\begin{figure*}
\begin{minipage}[b]{0.64\linewidth}
  \centering
  \includegraphics{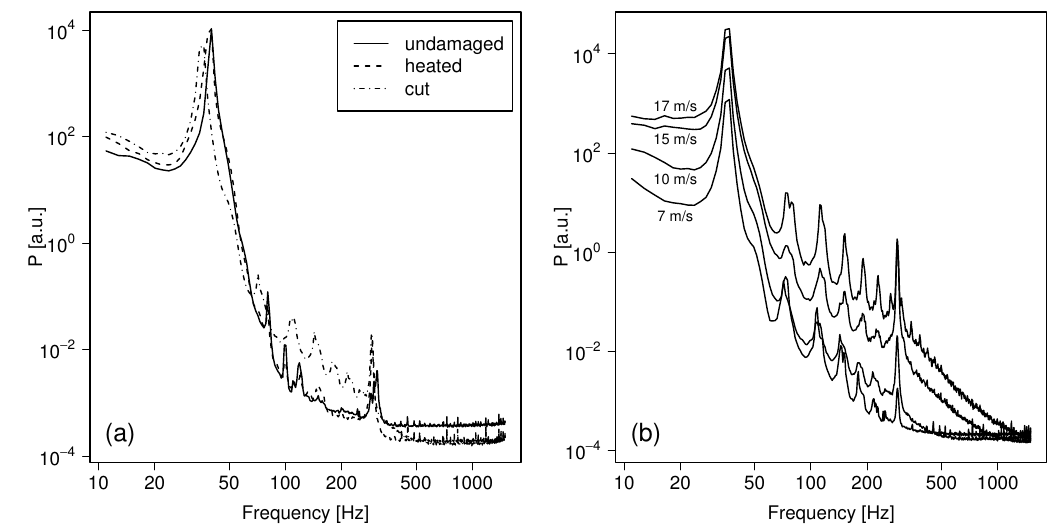}
  \caption{\label{spectrum} \label{spectrum_speeds} Power density spectra for the deflection signals in $y$-direction (a) at $\bar{u} = 10\,$m/s for different damages and (b) of the cut beam for different wind speeds.}
\end{minipage}
\hspace{0.3cm}
\begin{minipage}[b]{0.32\linewidth}
  \centering
  \includegraphics{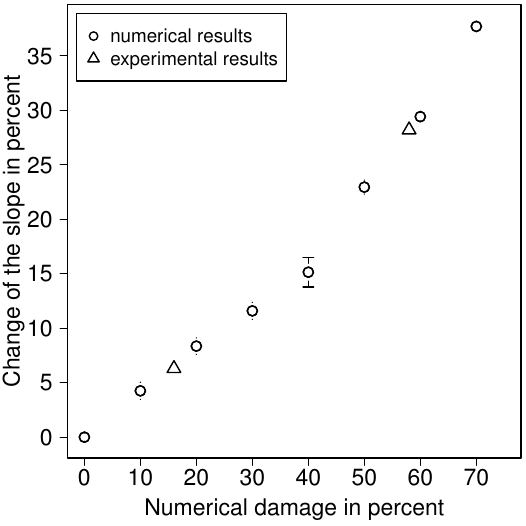}
  \caption{\label{diff_slope_FEM} Change of the slope of the drift function for increasing damage (numerical simulation) marked by open circles. Corresponding damages of the experiments marked by open triangles.}
\end{minipage}
\end{figure*}

As a next aspect we compare our results with the common damage detection method using power density spectra. Figure~\ref{spectrum}a shows the power density spectra of the deflection in $y$-direction for the different beams. The first eigenfrequency of the beam structure is very pronounced and the peaks are quite broad due to the noisy excitation. The shift in the first eigenfrequency for the heated beam is only $0.6\,$\% which is almost undetectable due to the broadened peak. The shift in frequency for the cut beam can be detected more clearly, it is $13.5\,$\%. Compared to the relative changes in the slope of $D^{(1)}$ of several percent for the heated and around $30\,$\% for the damaged beam the frequency analysis is less sensitive.

A comparison with power density spectra for different wind speeds shows that with increasing wind speed more and more eigenfrequencies get excited (cf. Fig.~\ref{spectrum_speeds}b). Taking also Fig.~\ref{spectrum}a into account, one can see that a damage excites higher eigenfrequencies in a similar way. (Additionally we realized that higher harmonics are excited too, if the measured signal saturates, e.g. by overloading the sensor or by bounded deflection of the structure.) The analysis of higher harmonics of the power spectra to quantify damages seems to become quite complicated under such conditions. From a point of view of system dynamics it is well known that damped relaxation systems like a beam structure may perform quite difficult nonlinear response dynamics under noisy excitations, which even may become chaotic. Consequently an analysis by a power spectrum of such systems is less appropriate than getting access to the underlying deterministic part of the response dynamics. We want to stress the point that our analysis by Kramers-Moyal coefficients is local in the stochastic variable $\bm{x}$ whereas power spectra are global in the sense that the full range of the phase space variable is processed.

At last, a first estimation of possible resolution of damages by our method is given. From simulations with systematic changes of the stiffness the changes of the slope of the deterministic part were evaluated as shown in Fig. \ref{diff_slope_FEM}. A damage of $5\,$\% is well resolvable. Putting the results from our experiments into relation of the numerical studies (see triangles in Fig. \ref{diff_slope_FEM}) the data shows that the obtained change in the slope of $6\,$\% (cf. Table~\ref{tabProzSlope}) corresponds to a (numerical) damage of about $15\,$\%. Thus, with our proposed method, it should be possible to detect even smaller damages of the mechanical structure than caused by our heating procedure.

\section{Discussion}
We showed that analyzing the deflection of a one-sided fixed beam structure by means of the reconstruction of the stochastic differential equation can be fruitful for structural health monitoring. The slope of the drift function is a sensitive indicator of the restoring force and thus of the mechanical properties. The sensitive detection of the drift function enables to show changes in the mechanical material properties and thus enables to detect probable damages. Most interestingly the method depends on noisy excitations caused by the environment. Noise helps to enlarge the phase space so that the conditional moments can be estimated properly. We also show evidence that the noise itself will be averaged out and will not have an influence on the absolute values of the deterministic drift functions. The ranges over which it can be reconstructed will be influenced by the noise (cf. Fig.~\ref{speeds}c). In this sense the method is robust against changing working conditions. We should point out that the deflections of our experimental set-up are in the range of $\mu$m, thus a linear response can be assumed. For larger deflections further corrections may have to be taken into account. Compared to the helpful influence of noise for the stochastic analysis, the effect of noise for the common analysis by determining the strength of excited eigenmodes as peaks in the power spectrum is more complicated and may even cause less sensitivity.

A very important application of this method are cases where one has no easy access to the considered system. We propose that this method can be used for remote diagnosis of running, embedded systems. Even changing working conditions are likely to be mapped onto the noise and thus will be filtered out by analyzing the drift function.
In principle it should be straightforward to apply the method to condition-monitoring of other systems and other noisy excitation forces, like for example parts of a running machinery.  For more complex dynamics like chaotic ones a higher dimensional phase space has to be used. There are even methods to verify by data analysis if such a higher dimensional analysis is required \cite{Siefert2003}.

\section{Acknowledgement}
The final version of this article is published in \textit{The European Physical Journal B} and is available at \href{www.epj.org}{www.epj.org}. DOI: \href{http://dx.doi.org/10.1140/epjb/e2012-30472-8}{10.1140/epjb/e2012-30472-8}.

\bibliography{paper}
\end{document}